\documentclass[12pt]{iopart}
\usepackage{graphicx}
\usepackage{braket}
\usepackage{subfig}

\begin{document}
\title{Narrow absorptive resonances in a four-level atomic system}
\author{M. G. Bason, A. K.  Mohapatra, K. J. Weatherill and C. S. Adams}
\address{Department Physics, Durham University, Rochester Building, South Road, Durham DH1 3LE, UK.}
\ead{m.g.bason@durham.ac.uk}
\begin{abstract}
We study the effect of a control beam on a $\Lambda$ electromagnetically induced transparency (EIT) system in $^{87}$Rb. The control beam couples one ground state to another excited state forming a four level $\mathcal{N}$--system. Phase  coherent beams to drive the $\mathcal{N}$--system are produced using a double injection scheme. We show that the control beam can be used to Stark shift or split the EIT resonance. Finally, we show that the when the control beam is on-resonance one observes a Doppler-free and sub-natural absorptive resonance with a width of order 100~kHz.  Crucially, this narrow absorptive resonance only occurs when atoms with a range of velocities are present, as is the case in a room temperature vapour. 
\end{abstract}
The interplay between two driving fields and a 3-level atom provides the simplest system for quantum interference phenomena \cite{Fleischauer2005}.  The understanding of coherent atom-field interactions has facilitated such advances as the slowing and storing of light \cite{Hau1999,Lukin2000,Phillips2001}, lasing without inversion \cite{Scully1994} and magnetometery \cite{Shah:2007il}. Common to all these is the process of electromagnetically induced transparency (EIT) \cite{Harris1997}, the signature of which is a narrow region of reduced absorption around the resonant frequency of a probe field.  Such spectral features, often called dark resonances, arise from the superposition of atomic states and driving fields.
This coherent non--linearity can lead to enhanced electro-optic effects \cite{Moha2008} with possible applications in photonic phase gates \cite{Friedler2005}.

More complicated systems of EIT are possible when one adds further driving fields and/or atomic states. For example, switching between subluminal and supraluminal light propagation has been studied in three level systems using a third driving field \cite{Agarwal2001,Sun2005}. Further to this, when four atomic states are coupled by three driving fields a double dark state results \cite{Yelin2003,Ye2002}. Most work on four level systems has concentrated on three ground terms and one excited state. Such systems have been shown to exhibit enhanced Kerr non-linearities \cite{Niu:05} and sub to superluminal light switching \cite{Cui2007}. Another type of four level system is the $\mathcal{N}$--system where there are two ground and two excited levels. This system is of particular interest due to the strongly enhanced ac Kerr effect \cite{Schmidt:96}. In addition, electromagnetically induced absorption (EIA) has been theoretically treated in such a system \cite{Goren2004} and experimental observation of a system using the D1 and D2 transitions of Rubidium have been made \cite{Kong:2007}.

Recently, we demonstrated how by coupling a $\Lambda$ system EIT to a Rydberg state, one can enable electro-optic control of the ground state coherence \cite{bason:032305}.
In this paper, we report the experimental observation and theoretical origin of narrow $\mathcal{N}$--system resonances in an atomic $^{87}$Rb vapour at room temperature. 
Starting with the `signature' peak in transmission associated with EIT, a third control field is introduced which couples the system to a fourth state. When the control field is off resonance we observe the ac Stark shift or Autler--Townes splitting of the EIT resonance. However, when the control field is on-resonance we observe that the EIT peak inverts, i.e., the medium is switched from transmitting to absorbing. Surprisingly, this absorptive resonance is both sub-Doppler and sub-natural, despite the fact that the final coupling is to a short lived excited state. Although similar features have been observed previously \cite{Ye, Wei:2007}, the sub-natural character of such resonances was not shown; we estimate a width of over 20 MHz in the latter work. We show that this sub-natural character of the $\mathcal{N}$--system resonance arises due to the contribution of off-resonant velocity classes in the thermal vapour. For these velocity classes the doppler shift experienced by an atom is greater than the decay rate of the excited state. Thus the effect is not predicted for cold atomic ensembles where the atomic motion is negligible. The enhanced absorption effect is demonstrated even when the Zeeman degeneracy of the ground state is lifted, thus the effect is distinct from EIA due to transfer of coherence \cite{Lezama1998} or transfer of population \cite{Goren2003} which arises due to degenerate levels. The ability to switch between narrow transparent and absorptive resonance or between sub-- and supra--luminal propagation is of interest in applications such as controlled pulse propagation and photon storage.

\section{Theory}

\begin{figure}[htbp]
\begin{center}
\includegraphics[width=0.4\linewidth]{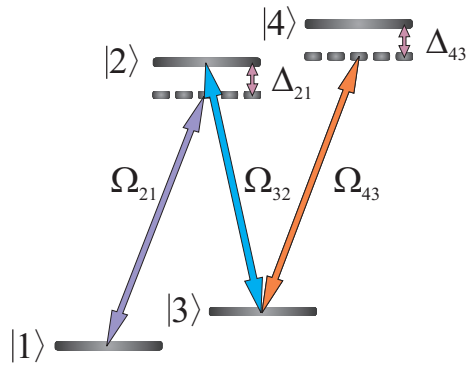}
\caption{Energy level diagram for the $\mathcal{N}$--system. Three fields with Rabi frequency  $\Omega_{ij}$ drive a 4-level system. The detunings from the excited states are given by $\Delta_{ij}$. We refer to the laser beams resonant with transitions $\vert 1\rangle \rightarrow \vert 2\rangle$, $\vert 2\rangle \rightarrow \vert 3\rangle$, and $\vert 3\rangle \rightarrow \vert 4\rangle$ as the probe, coupling and control beams, respectively.}
\label{Nleveldiagram}
\end{center}
\end{figure}

We consider a 4-level atomic system consisting of a typical $\Lambda$ system of two hyperfine ground terms, $|1\rangle$ and $|3\rangle$, and an excited third level, $|2\rangle$. An additional excited state level, $|4\rangle$, creates the $\mathcal{N}$-- type system shown in figure \ref{Nleveldiagram}. The four levels are coupled by probe, coupling and control laser fields with Rabi frequencies $\Omega_{21}$, $\Omega_{32}$ and $\Omega_{43}$, respectively. In the weak probe limit \cite{Siddons2008},  we solve the optical Bloch equations for the four level atom with co-propagating  $\Omega_{21}$ and $\Omega_{32}$ and counter propagating $\Omega_{43}$ fields. The Hamiltonian for this 4-level system is \begin{eqnarray}
H=\frac{1}{2}\left(
\begin{array}{cccc}
  \Delta_{21}-kv & \Omega_{21}  & 0  &0\\
  \Omega_{21}   & \Delta_{32}-kv  & \Omega_{32} &0 \\
  0&   \Omega_{32} & 0 & \Omega_{43}\\
  0& 0&\Omega_{43}&\Delta_{43}+kv
\end{array}
\right)
\label{Hamiltonian}
\end{eqnarray}
with atomic velocity $v$ and detunings $\Delta_{ij}$ as in figure \ref{Nleveldiagram}. The wavevector $k$, has the same magnitude for all fields. The decay terms are given by $\gamma_{ij}=(\Gamma_i+\Gamma_j)/2$, where $\Gamma_i$ is the natural decay rate of state $\ket{i}$. In this system $\Gamma_{2}=\Gamma_{4}=\Gamma=(2\pi{}\times)$ 6~MHz and $\gamma_{33}$ is taken as (2$\pi{}\times$) 100~kHz, to account for the transit time broadening. This Hamiltonian is then used to find the density matrix, $\sigma$, by solving the Liouville equation,
\begin{equation}
\dot{\sigma{}}=\frac{1}{i\hbar}[H, \sigma] - \gamma{}\sigma
\label{Master}
\end{equation}
in the steady state. Relaxation processes are included using $\gamma$, the decay matrix. The imaginary part of the coherence between $\ket{1}$ and $\ket{2}$ corresponds to the absorption of the probe beam.
\begin{figure}[h]
\begin{center}
\includegraphics[width=0.75\linewidth]{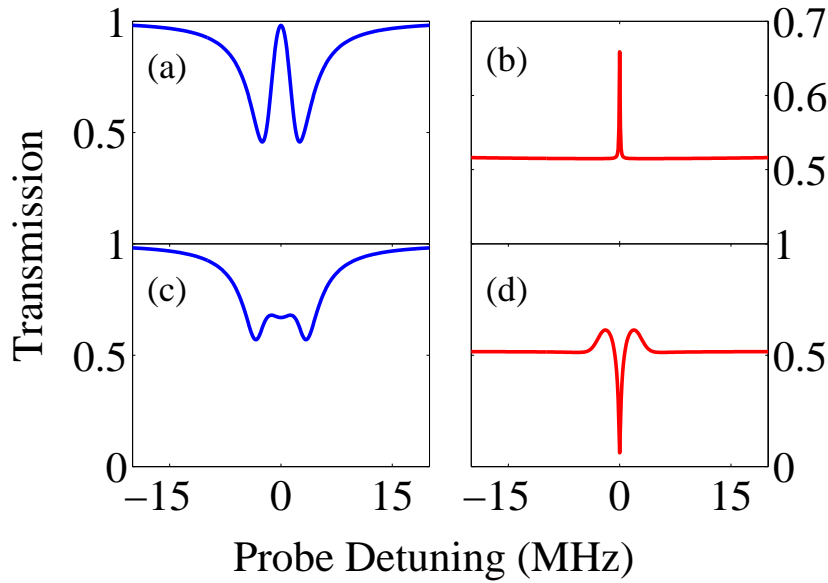}
\caption{Probe transmission as a function of detuning for both stationary atoms (a, c) and a room temperature ensemble with a Gaussian velocity distribution (b, d). For (a, c) the optical depth is chosen to match that of a 7.5~cm long room temperature Rb cell shown in  (b, d). In both cases $\Omega_{32}$=(2$\pi{}\times$) 5~MHz and states $\ket{2}$ and $\ket{4}$ have a decay of $\Gamma$=(2$\pi{}\times$) 6~MHz, while $\Gamma_{33}$ = (2$\pi{}\times$) 100 kHz. In (c) and (d) $\Omega_{43}$=(2$\pi{}\times$) 5~MHz. The probe and coupling beams co-propagate while the control beam counter-propagates.}

\label{DDK_theory}
\end{center}
\end{figure}
Figure \ref{DDK_theory} illustrates the difference between the absorption properties of cold atoms and a  thermal ensemble for the $\Lambda$ and $\mathcal{N}$ (4-level) systems. As previously noted \cite{Ye2002,Rostovtsev2002,pack:013801,Iftiquar:063807} a thermal ensemble has a narrower EIT resonance due to the contribution from off-resonant velocity classes, as seen in figures \ref{DDK_theory}(a) and (b). The application of the third laser field to a cold atom system only serves to broaden and split the transmission peak (figure \ref{DDK_theory}(c)), whereas for thermal atoms the resonance switches from transmitting to absorbing (figure \ref{DDK_theory}(d)). The largest absorption occurs when the Rabi frequencies of the control and coupling beams are equal. This condition gives rise to a Doppler--free eigenstate as discussed in \cite{Ye}. Crucially, the $\sim{}100$ kHz width of this pronounced absorptive resonance remains both sub-Doppler and sub-natural, despite the additional coupling to an excited state and the odd number of laser beams. This width is primarily limited by transit time broadening, whilst the height decreases for higher excited state decay rates. Limiting the number of velocity classes in the calculation shows the evolution of the EIT resonance into an absorptive feature when $kv\geq\Gamma$.
\section{Experimental Setup}
To observe the narrow resonance features expected in the EIT $\mathcal{N}$--system, it is necessary to make phase coherent laser beams separated in frequency by the ground state hyperfine splitting (6.8~GHz for $^{87}$Rb) \cite{Kasevich1991}. Several different approaches have previously been adopted to achieve this end. For instance, the injection locking of two diode lasers using the $\pm1$ diffraction orders from a high frequency acousto-optic modulator (AOM) \cite{Bouyer}, or injection locking of two diode lasers to the sidebands produced by current modulation of the master laser \cite{Matt}. Another approach is to use electro-optic modulation to produce sidebands at the required frequency before injection locking a slave laser. A drawback to this approach is that typically the spectral purity of the emerging beam is poor and requires filtering of the carrier frequency using a stabilised cavity \cite{Dotsenko2004}.

In this work we produce the required frequencies using electro--optic modulation followed by a double injection locking technique.  A schematic of the experimental setup is shown in figure 3 (a). A commercial 780.24 nm extended cavity diode laser is stabilised to the 5s~$^{2}\mathrm{S}_{1/2}(\mathit{F}=2)\rightarrow5$p$~^{2}\mathrm{P}_{3/2}(F'=2)$ transition in $^{87}$Rb ($\ket{2}\rightarrow{}\ket{3}$ in figure \ref{Nleveldiagram}) using polarisation spectroscopy \cite{Polspec}. A part of this light is separated using a beamsplitter and is used as the coupling beam.
\begin{figure}[ht]
	\begin{minipage}[b]{0.6\linewidth}
	\centering
\includegraphics[width=0.9\linewidth]{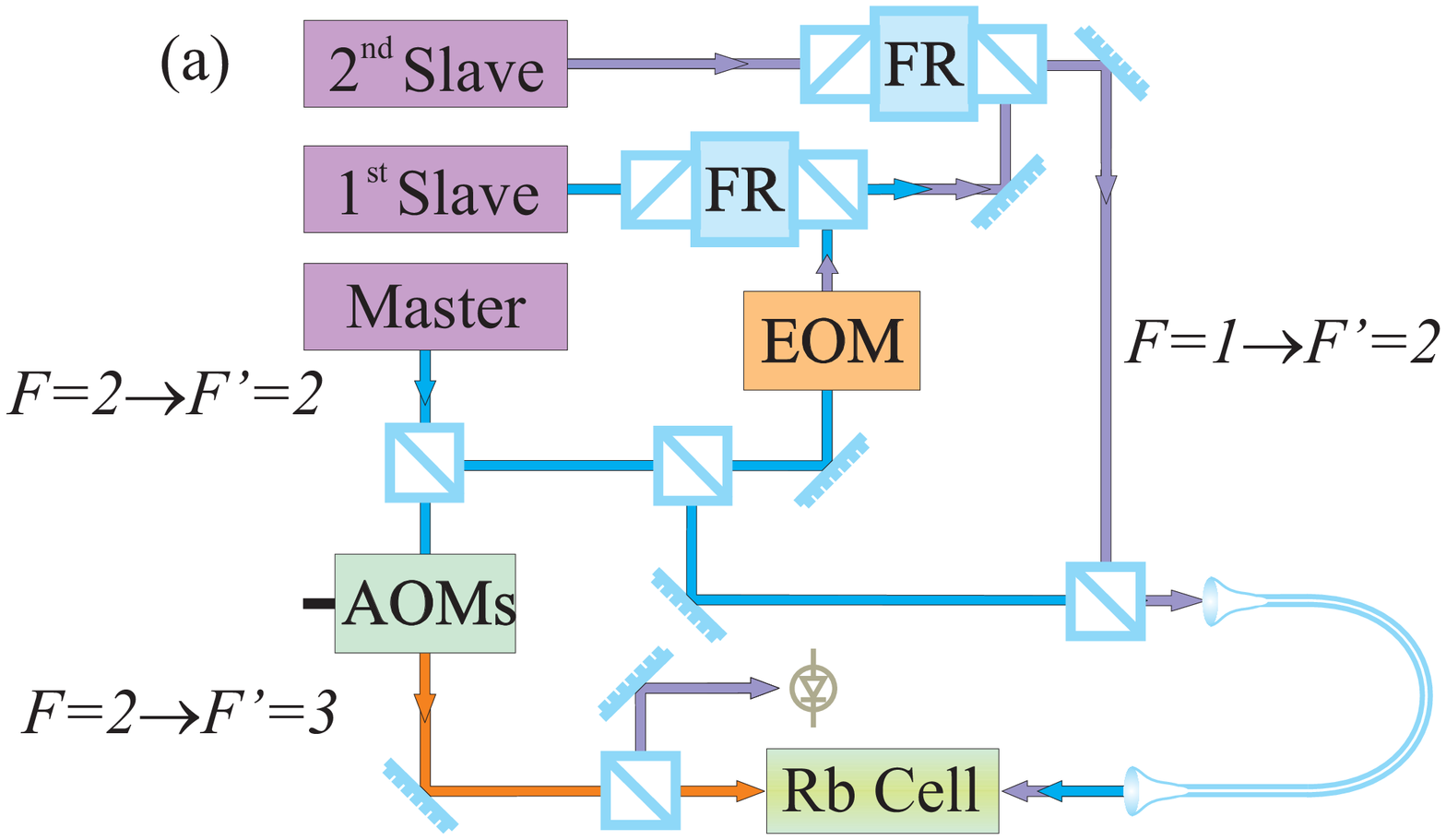}
\end{minipage}
\hspace{0.1cm}
\begin{minipage}[b]{0.4\linewidth}
\centering
\includegraphics[height=0.8\linewidth]{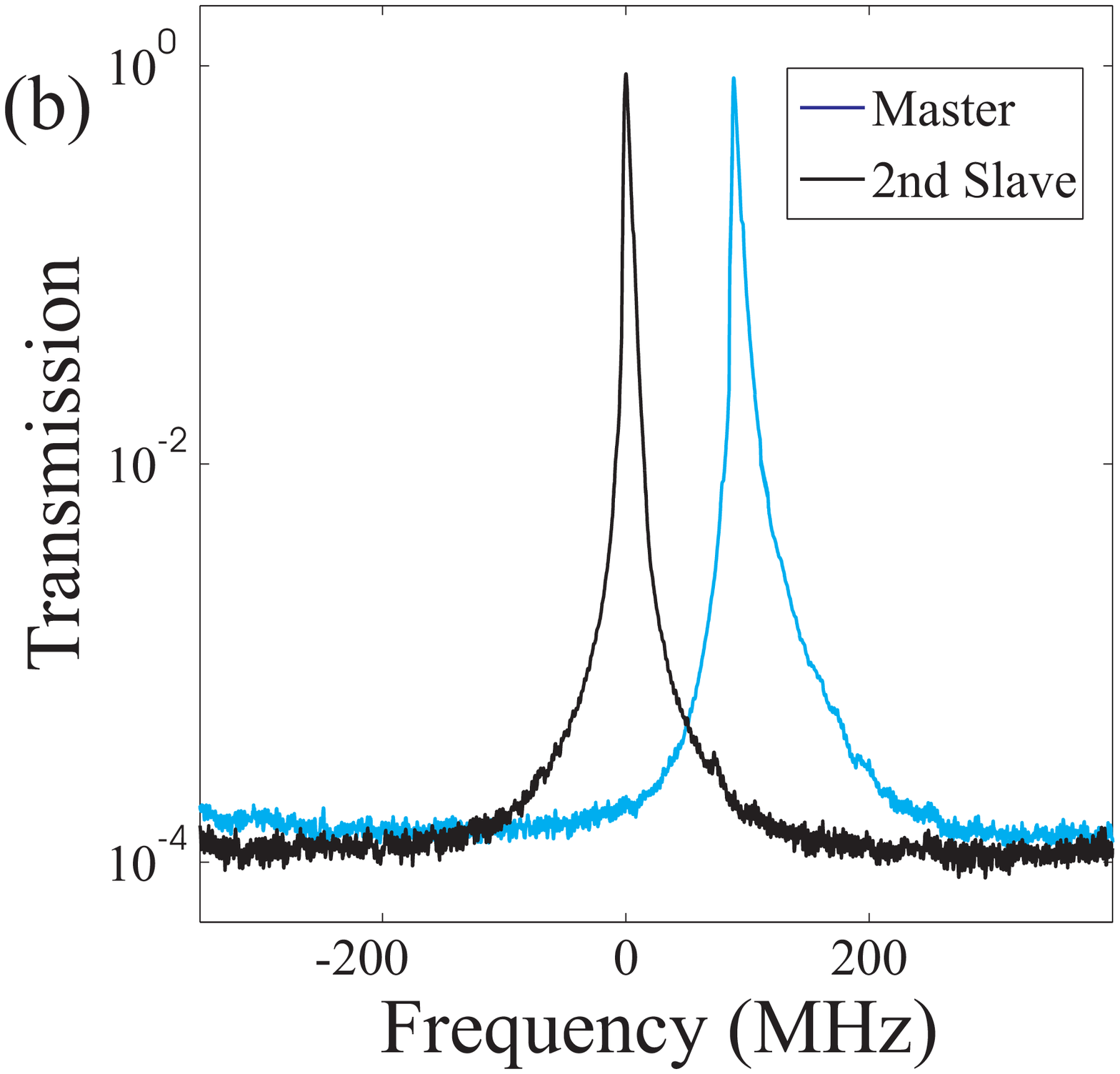}
\end{minipage}
\caption{(a) Schematic of the optical setup used to generate sidebands at 6.8 GHz.  A master laser is locked  on the $F=2\rightarrow{}F'=2$ resonance and the beam passes through an electro-optic modulator (EOM) at 6.8~GHz. This light then passes through a Faraday Rotator (FR) and injection locks a slave laser to the lower sideband. The light from this laser then seeds a second slave laser; the output of which is combined with the master light on a polarising beam splitter cube and coupled into a fiber. The third control beam frequency is derived from a double-passed acousto-optic modulator (AOM). This light counter-propagates with the master and slave light through the thermal vapour Rb cell. (b) The spectral purity of the second slave laser is observed by measuring the transmission through a scanning Fabry-Perot etalon with free spectral range $\sim$ 750~MHz. No component of master light on the second slave laser frequency spectrum is detected down to the noise level; a suppression of at least -40 dB.}
\label{40dB}
\end{figure}

The remainder of the light propagates through an electro-optic modulator (EOM) (New Focus 4851). The light frequency spectrum at this point is composed of the carrier frequency and two sidebands at $\pm$6.8~GHz.  Each sidebands has a relative intensity of 2\% of the carrier beam. Around 200 $\mathrm{\mu{}W}$ of this light is then used to seed a free running diode laser which locks to the lower sideband \cite{Szymaniec1997, Clarke1998}. This is optimised using a combination of current and temperature tuning, resulting in a purity of around -20~dB. To achieve higher spectral purity and greater stability, a second slave laser is injected with the output of the first.  The spectral output of this third laser, shown in figure \ref{40dB}(b) has carrier suppression to the level of at least -40 dB.  This light is then resonant with the 5s~$^{2}\mathrm{S}_{1/2}(\mathit{F}=2)\rightarrow5$p$~^{2}\mathrm{P}_{3/2}(F'=2)$ transition ($\ket{1}\rightarrow{}\ket{2}$ in figure \ref{Nleveldiagram}) and is used as the probe beam. For near resonant coupling to the 5s~$^{2}\mathrm{S}_{1/2}(\mathit{F}=2)\rightarrow5$p$~^{2}\mathrm{P}_{3/2}(F'=3)$ ($\ket{3}\rightarrow{}\ket{4}$ in figure \ref{Nleveldiagram}) transition two AOMs shift a fraction of the master light 267 MHz to the blue, to give the control beam.  For larger detunings the frequency requirements are less stringent, as the light is detuned further than the Doppler width. In this case another external cavity diode laser is used, which is not stabilised relative to the other lasers.

The probe and coupling beams are combined on a polarisation beam splitting cube and coupled into a polarisation maintaining single mode fiber. The output of the fiber is collimated and propagates through a magnetically shielded, Rb cell of length 75 mm as shown in figure \ref{40dB}(a). The probe and coupling beams have a waist of 1.3 mm (1/e$^{2}$ radius) and have orthogonal linear polarisations. The control beam counter-propagates with the other two beams, as in equation (\ref{Hamiltonian}). By changing the drive frequency of the EOM the probe beam can be scanned through the $F=1\rightarrow{}F' =2$ resonance. A polarising beam splitting cube is used to pick off the coupling beam, allowing the probe transmission to be monitored as a function of detuning.

\section{Results and Discussion}

\begin{figure}[h!]
\begin{minipage}[b]{0.3\linewidth}
\centering
\includegraphics[width=0.9\linewidth]{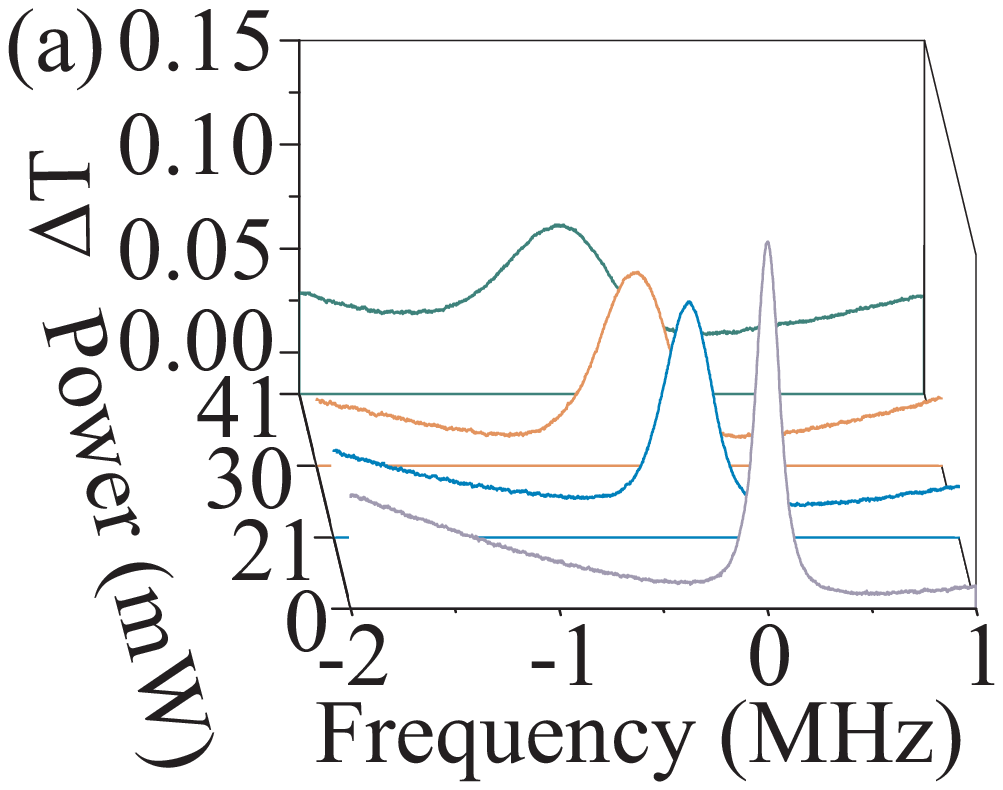}
\end{minipage}
\hspace{0.1cm}
\begin{minipage}[b]{0.3\linewidth}
\centering
\includegraphics[width=0.9\linewidth]{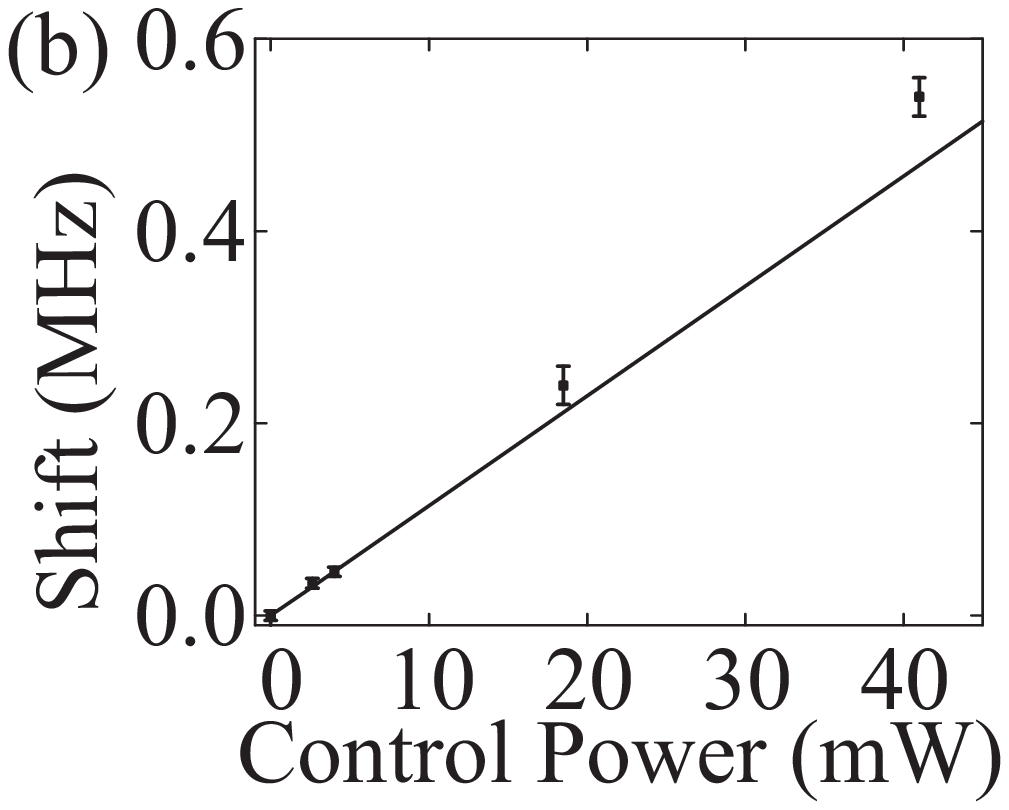}
\end{minipage}
\hspace{0.1cm}
\begin{minipage}[b]{0.3\linewidth}
\centering
\includegraphics[width=0.9\linewidth]{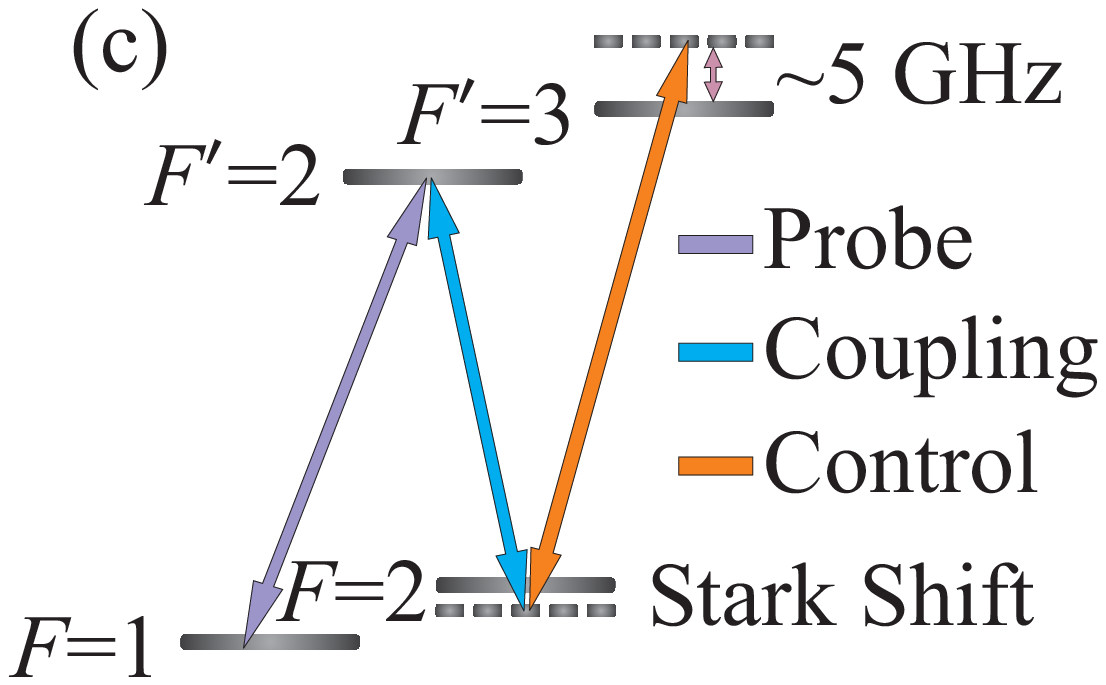}
\end{minipage}
\caption{ (a) Change in transmission as a function of detuning from the $5s^2S_{1/2}(F=1)\rightarrow 5p ^{2}P_{3/2}(F'=2)$ level. (b) Shift in frequency of EIT peak for various control beam powers. Data points are taken from (a), whilst the solid line is the calculated ac stark shift. (c) Level scheme for the $\mathcal{N}$--system A far off-resonant control field is used to shift the $F=2$ ground state}
\label{stark_shifting_levels}
\end{figure}

Transmission spectra for the $\mathcal{N}$--system for an off--resonant control beam are shown in figure \ref{stark_shifting_levels}(a). Using a probe power of 20 $\mathrm{\mu{}W}$, coupling power of  110 $\mathrm{\mu{}W}$ and no control beam a narrow EIT signal of width $\sim{}$70 kHz is observed. This width is limited by transit--time broadening. When a control beam is added, detuned below resonance by approximately 5~GHz as shown in figure \ref{stark_shifting_levels}(c), the resulting spectra show that the EIT feature broadens and shifts to lower frequency. This shift can be understood as an ac Stark shift of the ground state, $\ket{3}$. In figure \ref{stark_shifting_levels}(b) the measured shift is shown relative to the predicted ac Stark shift: $\Delta{}E=\pm{}\hbar{}\Omega^2/4\Delta$, where $\Omega$ and $\Delta$ are the control beam Rabi frequency and detuning respectively.
The broadening of the EIT peak is expected as the shift experienced by each atom in the ensemble is dependent upon its position relative to the spatially inhomogeneous intensity of the control laser field. 

Similar characteristics are seen when the Zeeman degeneracy is lifted by applying $\sim$1~G axial magnetic field. In this case three different EIT peaks corresponding to EIT $\Lambda$ systems consisting of different $m_F$ states are observed, see figure 5(a). Figure 5(b) shows the spectra for the $\mathcal{N}$--system with the 110 $\mathrm{\mu{}W}$ control beam detuned by -5~MHz. In this small detuning limit the effect of the control beam is to Autler-Townes split the EIT resonance \cite{yang2005} rather that just shift it as in figure 4(a). The results in figures 5(a) and (b) can be understood by considering the couplings for $\sigma^+$ $\sigma^+$ transitions in figure 5(c). For the $\Lambda$ system, due to the beam wavevectors being in the same direction as the magnetic field, no $\pi$ transitions are expected. This configuration gives resonances at three frequencies $\alpha$, $\beta$ and $\chi$; the case of $\sigma^-$ $\sigma^-$ beams is identical due to symmetry. For mixed transitions, $\sigma^+$ $\sigma^-$ and $\sigma^-$ $\sigma^+$, the resonances also occur at these three probe frequencies. One example is $\ket{2,-1}\rightarrow{}\ket{2,0}\rightarrow{}\ket{1,1}$ which occurs at the same probe frequency as the $\beta$ system. When the control laser field is applied to the ensemble to create the $\mathcal{N}$--system, an asymmetric Autler--Townes splitting of the EIT feature is observed \cite{yang2005}. The relative height of the components are determined by optical pumping effects. As in figure 4a, the strongly shifted peak is also broadened due to the inhomogeneous intensity profile of the control beam.
\begin{figure}[h]
	\begin{minipage}[h]{0.5\linewidth}
	\centering
\includegraphics[width=0.9\linewidth]{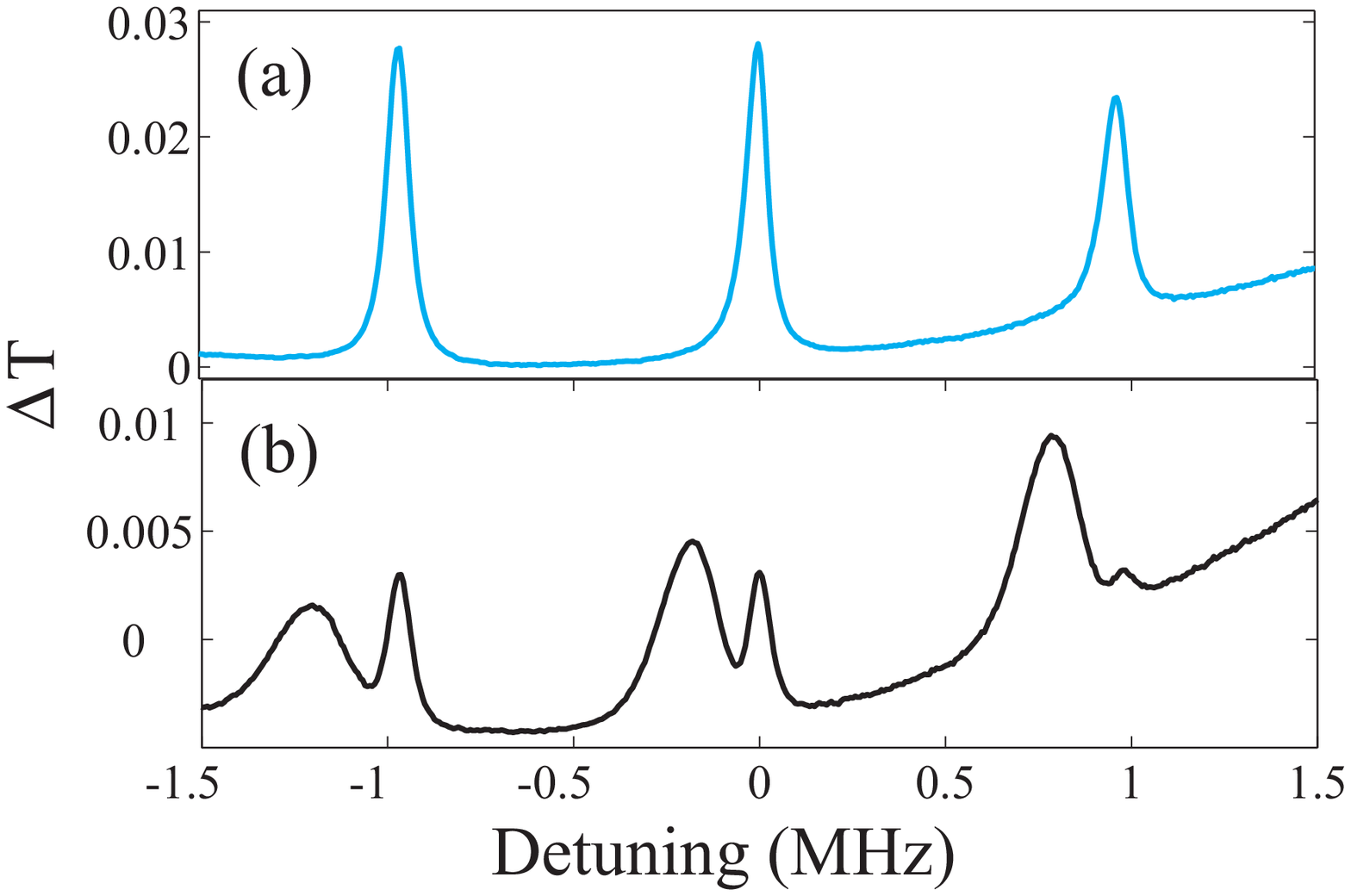}
\end{minipage}
\hspace{0.1cm}
\begin{minipage}[h]{0.5\linewidth}
\centering
\includegraphics[height=0.6\linewidth]{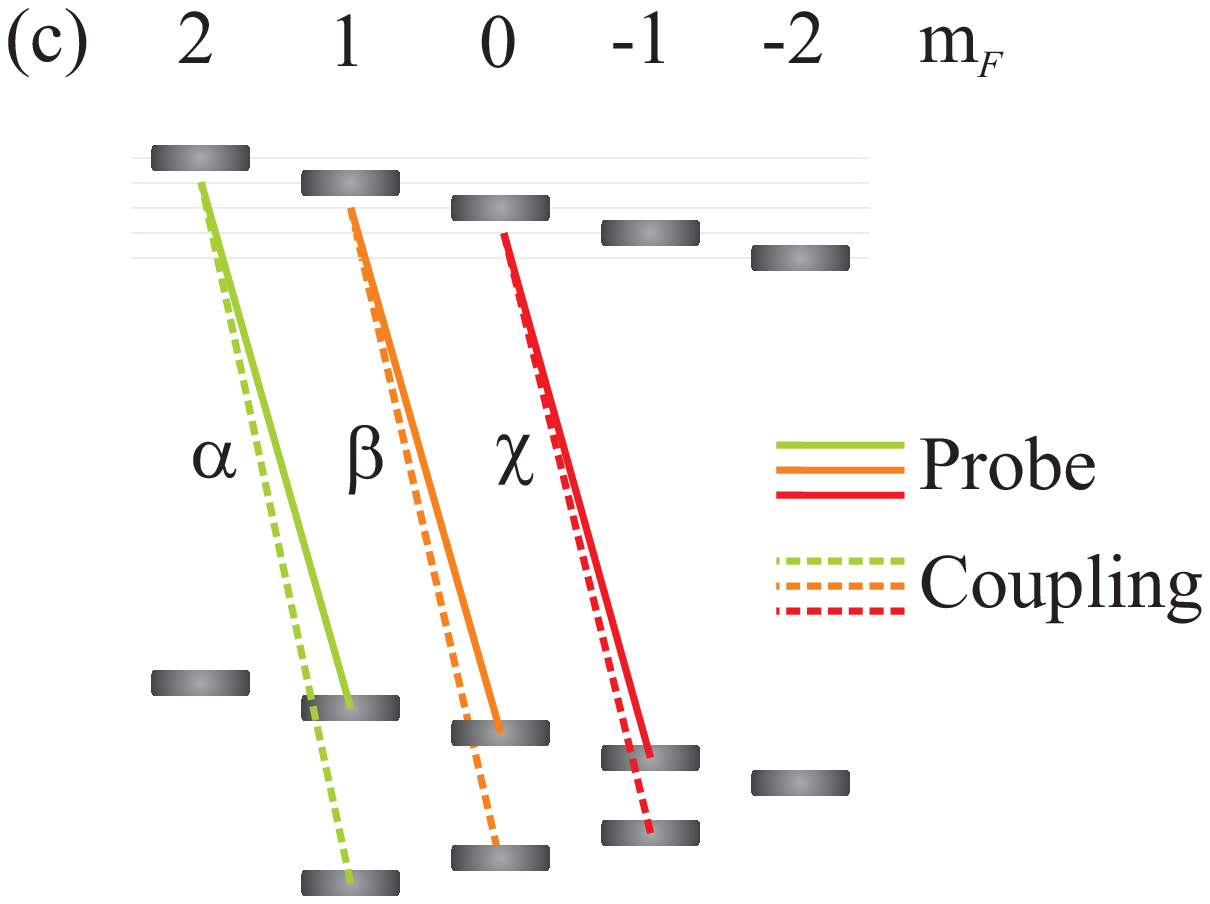}
\end{minipage}
\caption{Change in probe transmission as a function of detuning from the  $F=1\rightarrow{}F'=2$ level, with a small magnetic field of order 1 Gauss to lift the degeneracy in the $F=1$ and $F=2$ ground states. In (a) three peaks in transmission are seen. (b) A control beam with 110 $\mathrm{\mu{}W}$, detuned by -5 MHz from the $F'=3$ level, is applied. The level scheme when only $\sigma{}^+$ transitions are present is shown in (c), Raman transitions are observed at energies $\alpha$, $\beta$ and $\gamma$ due to the lifted Zeeman degeneracy. $\sigma{}^-$ transitions are omitted for clarity.}
\label{b_shifting_levels}
\end{figure}
\begin{figure}[h]

\subfloat{\label{fig:data}\includegraphics[width=0.3\linewidth]{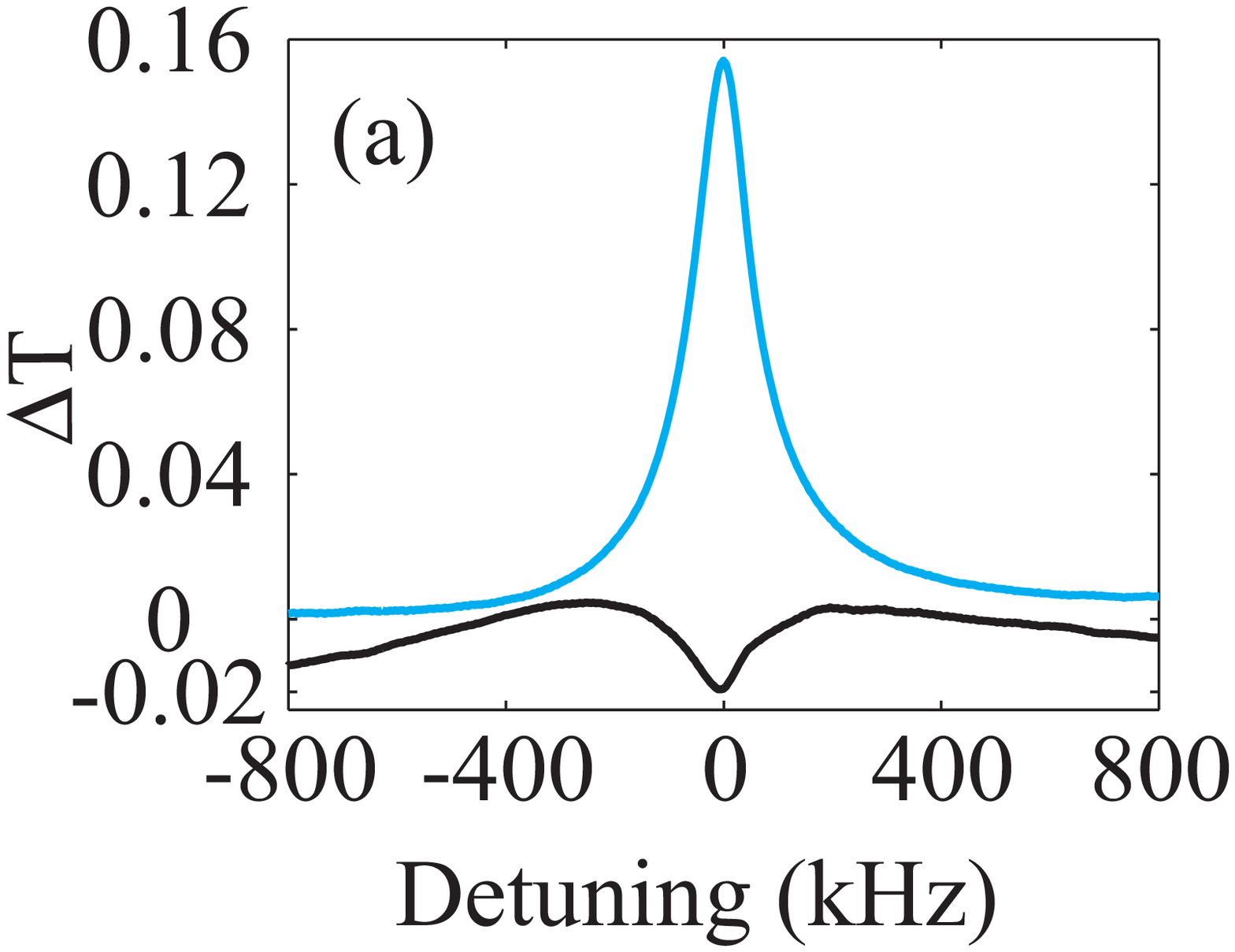}}
\subfloat{\label{fig:data2}\includegraphics[width=0.3\textwidth]{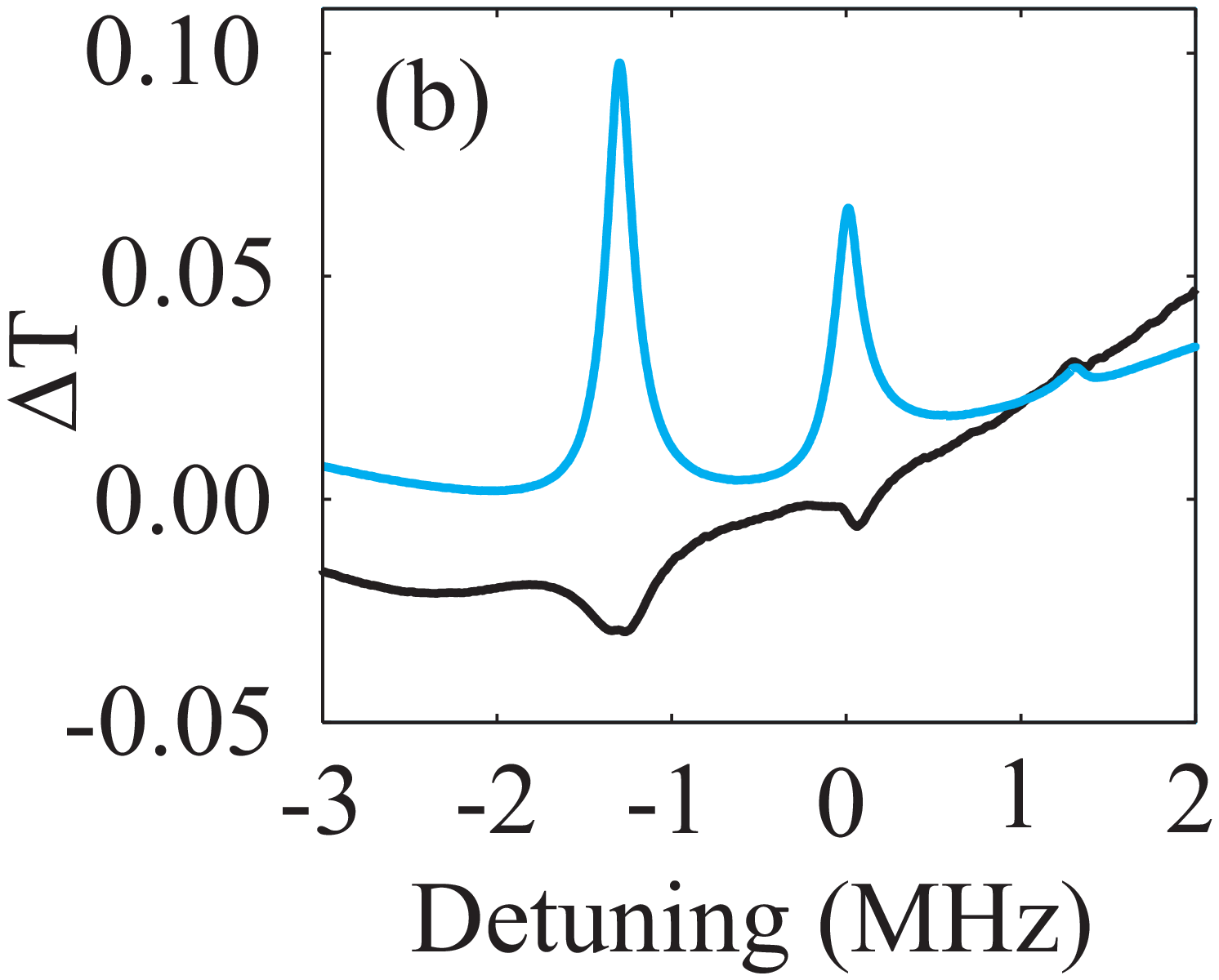}}
\subfloat{\label{fig:pic}\includegraphics[width=0.3\textwidth]{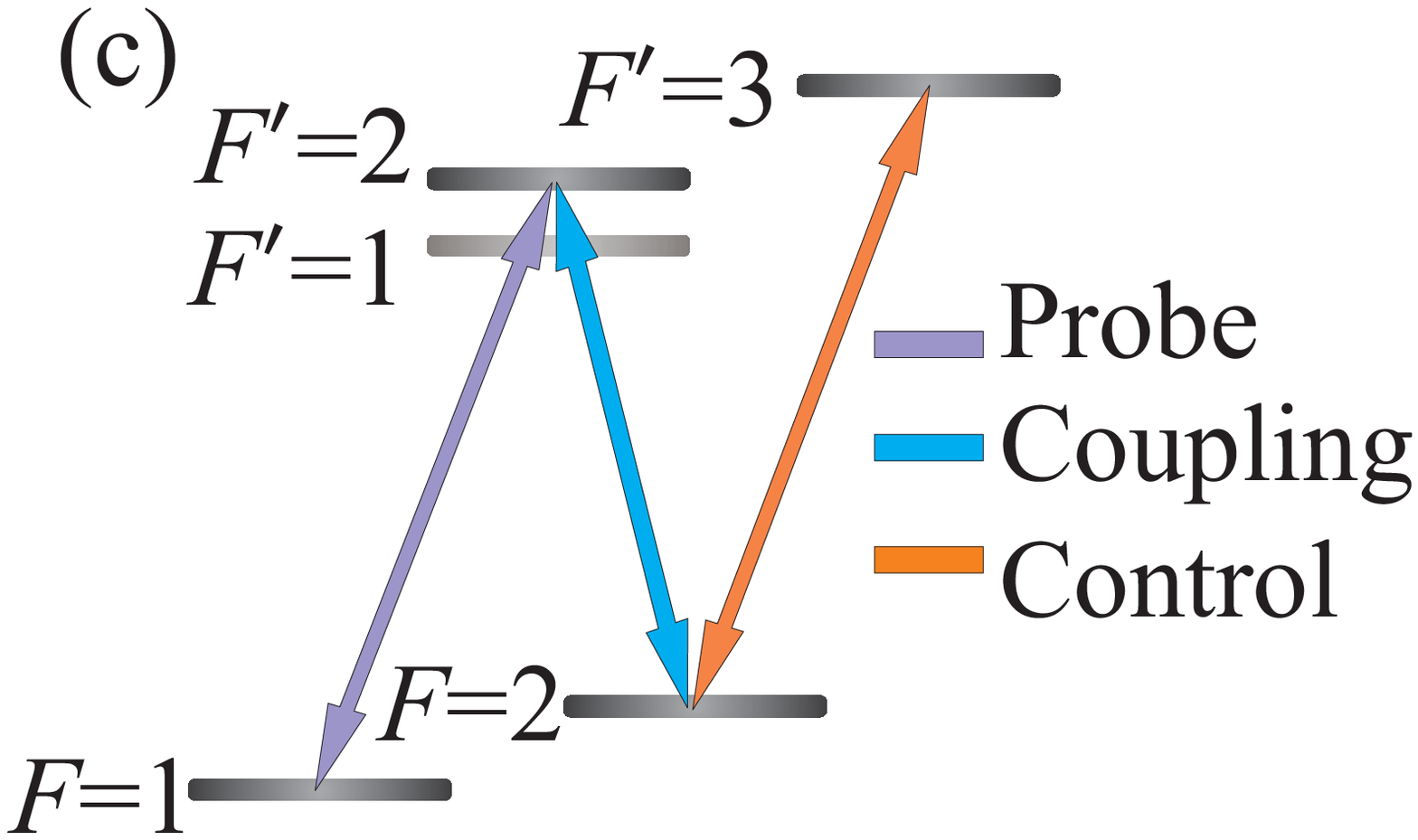}}
\caption{(a) Change in probe transmission as a function of detuning from the\\$\mathit{F}=1\rightarrow{}F'=2$ level, coupling beam on resonance, for no control beam (blue) and control beam on resonance (black). (b) Despite the lifted Zeeman degeneracy using a $\sim1$ Gauss field, the absorptive resonance is still seen. The relative peak heights are determined by optical pumping. (c) Level scheme for observation of sub-natural enhanced absorption. Both the coupling and control lasers are on resonance, and the probe laser is scanned through resonance.}
\label{4levels3}

\end{figure}

Finally, we consider the case of the control beam tuned to resonance as shown in figure \ref{4levels3}(c). Here the narrow transparency peak is observed to switch to an absorptive resonance with sub--natural linewidth as seen in figure \ref{4levels3}(a).
This result is in qualitative agreement with the theoretical predictions presented in section 1. However, the enhanced absorption is smaller than that predicted in figure \ref{DDK_theory} due to multiple contributions from the 5p$~^{2}\mathrm{P}_{3/2}$ hyperfine states, i.e., the system is not composed of only 4 states as was assumed in the theory section. The same effect, i.e. a switching between transparency and absorption is also observed when the Zeeman degeneracy is lifted, figure \ref{4levels3}(b). This demonstrates that the enhanced absorption observed in the $\mathcal{N}$--system has a different origin to the EIA effect occurring in degenerate systems \cite{Goren2004}.

There are a number of additional interesting features of this result. In particular the enhanced absorption resonance clearly has a sub-natural linewidth (of order 100~kHz) even though it is induced by coupling to a state with a width of 6~MHz. As shown in the theory section this arises due to the contribution of off resonant velocity classes. The technique could be employed to enhance the optical depth of an otherwise optically thin sample with potential applications in detecting weak absorption lines \cite{moha2007} or light storage. Future work will focus on the D1 line where the larger hyperfine splitting in the excited state (362~MHz) means that the larger absorption predicted by the model (figure 2) can be observed.

\section{Summary}

In summary, we present a novel scheme to create phase coherent laser beams separated in frequency by the Rb hyperfine ground state splitting by double injection locking. Using this set up we  demonstrate a narrow absorptive resonance in a 4 level $\mathcal{N}$--system. This resonance is both Doppler-free and sub-natural despite the coupling to an excited state. This behaviour is shown to be a feature of the thermal atomic system and is predicted not to occur for cold atoms.

\begin{ack}
We thank I. G. Hughes and M. P. A. Jones for valuable discussions, S. J. Smith for help with theoretical modelling and EPSRC for financial support.  
\end{ack}

\bibliographystyle{nature} 
\bibliography{DDK-2} 
\end{document}